\documentclass[aip,cha,twocolumn,showpacs,preprintnumbers,amsmath,amssymb,superscriptaddress,10pt]{revtex4-1}

\usepackage{amsmath,amssymb}
\usepackage{bm}
\usepackage[latin1]{inputenc}
\usepackage{dcolumn}
\usepackage{graphicx}
\usepackage{SIunits}
\usepackage{multirow}
\usepackage{ae}
\usepackage{subfigure}

\def\plotwidth{8cm}

\begin{document}

\title{\Large Hacking commercial quantum cryptography systems by tailored bright illumination}

\author{Lars Lydersen}
\email{lars.lydersen@iet.ntnu.no}
\affiliation{Department of Electronics and Telecommunications, Norwegian University of Science and Technology, NO-7491 Trondheim, Norway}
\affiliation{University Graduate Center, NO-2027 Kjeller, Norway}

\author{Carlos Wiechers}
\affiliation{Max Planck Institute for the Science of Light, G\"{u}nther-Scharowsky-Str. 1/Bau 24, 91058 Erlangen, Germany}
\affiliation{Institut f\"{u}r Optik, Information und Photonik, University of Erlangen-Nuremberg, Staudtstra\ss e 7/B2, 91058, Erlangen, Germany}
\affiliation{Departamento de F\'{i}sica, Universidad de Guanajuato, Lomas del Bosque 103, Fraccionamiento Lomas del Campestre, 37150, Le\'{o}n, Guanajuato, M\'{e}xico}

\author{Christoffer Wittmann}
\author{Dominique Elser}
\affiliation{Max Planck Institute for the Science of Light, G\"{u}nther-Scharowsky-Str. 1/Bau 24, 91058 Erlangen, Germany}
\affiliation{Institut f\"{u}r Optik, Information und Photonik, University of Erlangen-Nuremberg, Staudtstra\ss e 7/B2, 91058, Erlangen, Germany}

\author{Johannes Skaar}
\affiliation{Department of Electronics and Telecommunications, Norwegian University of Science and Technology, NO-7491 Trondheim, Norway}
\affiliation{University Graduate Center, NO-2027 Kjeller, Norway}

\author{Vadim Makarov}
\affiliation{Department of Electronics and Telecommunications, Norwegian University of Science and Technology, NO-7491 Trondheim, Norway}

\date{9 July 2010}

\maketitle

\textbf{The peculiar properties of quantum mechanics allow two remote parties to communicate a private, secret key, which is protected from eavesdropping by the laws of physics \cite{mayers1996,lo1999,shor2000,bennett1984}. So-called quantum key distribution (QKD) implementations always rely on detectors to measure the relevant quantum property of single photons \cite{scarani2009}. Here we demonstrate experimentally that the detectors in two commercially available QKD systems can be fully remote-controlled using specially tailored bright illumination. This makes it possible to tracelessly acquire the full secret key; we propose an eavesdropping apparatus built of off-the-shelf components. The loophole is likely to be present in most QKD systems using avalanche photodiodes to detect single photons. We believe that our findings are crucial for strengthening the security of practical QKD, by identifying and patching technological deficiencies.}

The field of quantum key distribution has evolved rapidly in the last decades. Today QKD implementations in laboratories can generate key over fibre channels with lengths up to $250\,\kilo\meter$~\cite{stucki2009} and a few QKD systems are even commercially available promising enhanced security for data communication. 

In all proofs for the security of QKD there are assumptions on the devices involved. However the components used for the experimental realizations of QKD deviate from the models in the security proofs. This has led to iterations where security threats caused by deviations have been discovered, and the loopholes have been closed either by modification of the implementation, or more general security proofs \cite{gottesman2004,fung2009,lydersen2010}. In other cases, information leaking to the eavesdropper was quantified~\cite{lamas-linares2007, nauerth2009}. 

Attacks exploiting the most severe loopholes are usually experimentally unfeasible with current technology. A prominent example is the photon number splitting attack~\cite{lutkenhaus2000} which requires the eavesdropper Eve to perform a quantum non-demolition measurement of the photon number sent by Alice. The attack is still unfeasible, and has been outruled by improved QKD protocols~\cite{hwang2003,scarani2004}. In contrast, a more implementation-friendly attack is the time-shift attack \cite{qi2007} based on detector efficiency mismatch \cite{makarov2006}. Experimentally however, this attack has only given a small information-theoretical advantage for Eve when applied to a modified version of a commercial QKD system \cite{zhao2008}. In the attack Eve captured partial information about the key in 4\% of her attempts, such that she could improve her random (brute-force) search over all possible keys. 

In this Letter, we demonstrate how two commercial QKD systems id3110 Clavis2 and QPN 5505, from the commercial vendors ID Quantique and MagiQ Technologies, can be fully cracked. We show experimentally that Eve can blind the gated detectors in the QKD systems using bright illumination, thereby converting them into classical, linear detectors. The detectors are then fully controlled by classical laser pulses superimposed over the bright continuous-wave (CW) illumination. Remarkably the detectors exactly measure what is dictated by Eve; with matching measurement bases Bob detects exactly the bit value sent by Eve, while with incompatible bases the bit is undetected by Bob. Even the detectors dark counts are completely eliminated (but can be simulated at will by Eve). Based on these experimental results we propose in detail how Eve can attack the systems with off-the-shelf components, obtaining a perfect copy of the raw key without leaving any trace of her presence. 

\begin{figure}[htbb]
  \subfigure[]{\includegraphics[width=8cm]{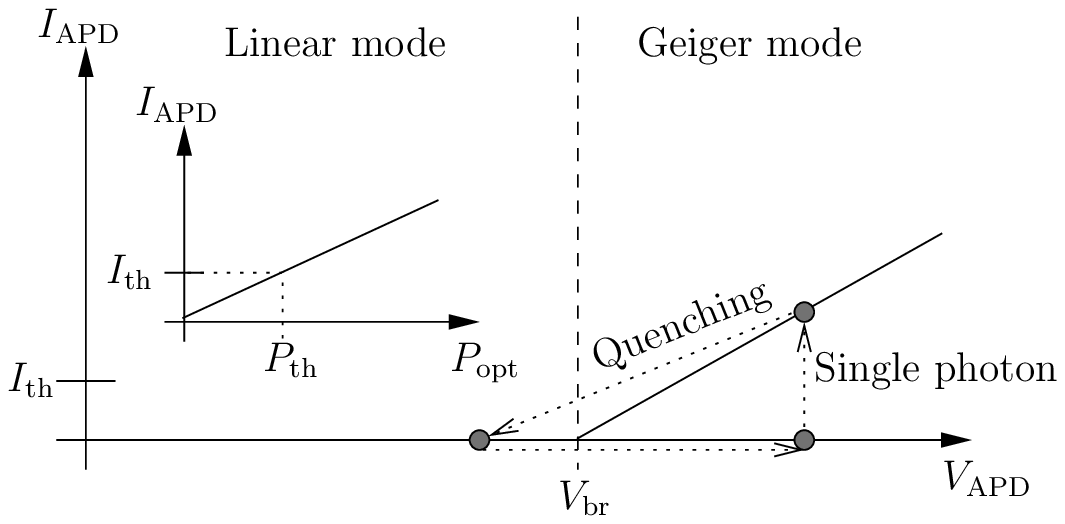}}
  \subfigure[]{\includegraphics[height=3.1cm]{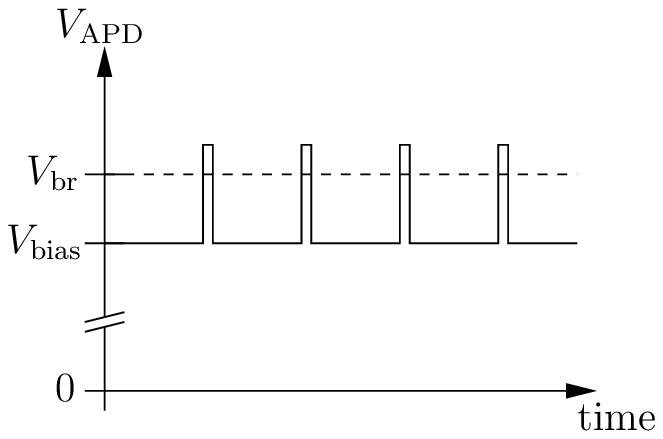}}
  \subfigure[]{\includegraphics[height=3.1cm]{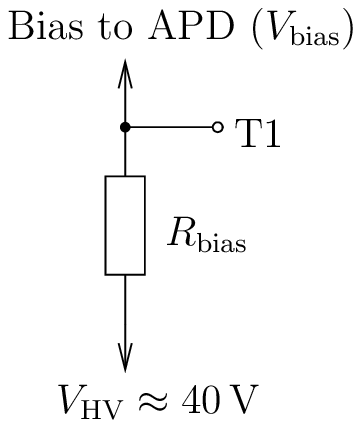}}
  \caption{APD as a single-photon detector. a) In Geiger mode where the APD is reverse-biased above the breakdown voltage $V_{\text{br}}$, an absorbed single photon causes a large current $I_{\text{APD}}$ through the APD. A detection signal called a \emph{click} occurs when $I_{\text{APD}}$ crosses the threshold $I_{\text{th}}$. Afterwards $V_{\text{APD}}$ is lowered below $V_{\text{br}}$ to quench the avalanche, before returning to Geiger mode. Below $V_{\text{br}}$, in the linear mode, the current $I_{\text{APD}}$ is proportional to the incident optical power $P_{\text{opt}}$. Then $I_{\text{th}}$ becomes an optical power threshold $P_{\text{th}}$. b) The commercial systems use gated detectors, with the APDs in Geiger mode only when a photon is expected, to reduce false detections called \emph{dark counts}. In practice, the APD is biased just below $V_{\text{br}}$, and periodical $\sim\!\!3\,\volt$ voltage pulses create Geiger mode time regions, so-called \emph{gates}. c) In both systems, the bias high-voltage supply $V_{\text{HV}}$ has impedance $R_{\text{bias}}$ ($R_{\text{bias}} = 1\,\kilo\ohm$ in Clavis2 and $20\,\kilo\ohm$ in QPN 5505) before $V_{\text{bias}}$ is applied to the APD at the point T1. Therefore, any current through $R_{\text{bias}}$ reduces $V_{\text{bias}}$ (see Supplementary information section~I for more details).}
  \label{fig:apds}
\end{figure}

Today most QKD systems use avalanche photo diodes (APDs) to detect single photons \cite{cova2004}. To detect single photons APDs are operated in Geiger mode (see Fig.~\ref{fig:apds}). 
\begin{figure}[htbp]
  \includegraphics[width=8cm]{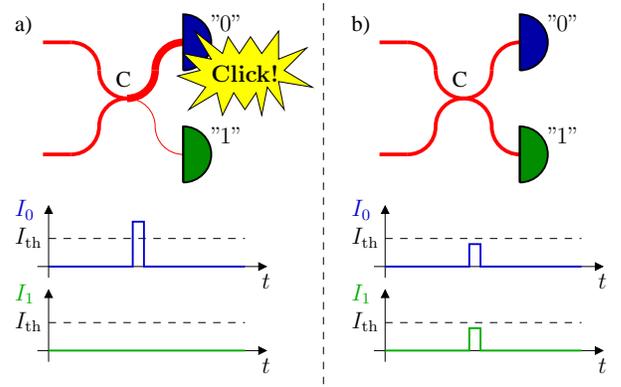}
  \caption{How Eve's trigger pulses are detected by Bob. Schemes show the last 50/50 coupler (C) and Bob's detectors in a phase-encoded QKD system. $I_0/I_1$ is the current running through APD 0/1. a) Eve and Bob have selected matching bases, and Eve detected the bit value 0. Therefore the trigger pulse from Eve interferes constructively and its full intensity hits detector~0. The current caused by Eve's pulse crosses the threshold current $I_{\text{th}}$ and causes a click. b) Eve and Bob have selected opposite bases. The trigger pulse from Eve does not interfere constructively and half of its intensity hits each detector. This causes no click as the current is below the threshold $I_{\text{th}}$ for each detector.}
  \label{fig:attack} 
\end{figure}
However all APDs spend part of the time being biased under the breakdown voltage, in the linear mode. During this time the detector remains sensitive to bright light with a classical optical power threshold $P_{\text{th}}$. If Eve has access to the APDs in the linear mode, she may eavesdrop on the QKD system with an intercept-resend (faked-state \cite{makarov2005,makarov2008a}) attack as follows: Eve uses a copy of Bob to detect the states from Alice in a random basis. Eve resends her detection results, but instead of sending pulses at the single photon level she sends bright trigger pulses, with peak power just above $P_{\text{th}}$. Bob will only have a detection event if his active basis choice coincides with Eve's basis choice (see Fig.~\ref{fig:attack}), otherwise no detector clicks. This causes half of the bits to be lost, but in practice this is not a problem because transmittance from the output of Alice to Bob's detectors is much lower than 1/2. Also Bob's APDs rarely have a quantum efficiency over 50\%, while the trigger pulses always cause clicks. For a Bob using passive basis choice, Eve launches the peak power just above $2P_{\text{th}}$ since half of the power hits the conjugate basis detectors \cite{makarov2008a,gerhardt}. Then Bob's detector always clicks. 

After the raw key exchange, Bob and Eve have identical bit values and basis choices. Since Alice and Bob communicate openly during sifting, error correction and privacy amplification, Eve simply listens to this classical communication and applies the same operations as Bob to obtain the identical final key. 

The attack is surprisingly general: all commercial QKD systems and the vast majority of research systems use APD-based detectors which all operate their APDs part time in linear mode. Detectors with passively- and actively-quenched APDs can also be kept in linear mode through blinding \cite{makarov2009,makarov2008a}. The attack works equally well on the Scarani-Acin-Ribordy-Gisin 2004 (SARG04) \cite{scarani2004} and decoy-state BB84 \cite{hwang2003} protocols as well as the normal BB84 protocol \cite{bennett1984}. With suitable modifications it applies to differential phase shift (DPS) \cite{takesue2005}, and given the right set of detector parameters to coherent one-way (COW) \cite{stucki2005} protocols. 

Note that the threshold $P_{\text{th}}$ should be sufficiently well defined for perfect eavesdropping. To be precise, let detector $i$ always click from a trigger pulse of optical peak power $\geq P_{\text{always},i}$, and never click from a trigger pulse of optical peak power $\leq P_{\text{never},i}$. The requirement for Eve to be able to make any single detector click while none of the other detectors click, can be expressed in terms of the click thresholds as 
\begin{equation}
  \max_i \left\{ P_{\text{always},i} \right\} < 2\left(\min_i \left\{ P_{\text{never},i} \right\}\right).
  \label{eq:attack-requirement}
\end{equation}

\begin{figure}[tbp]
  \includegraphics[width=\plotwidth]{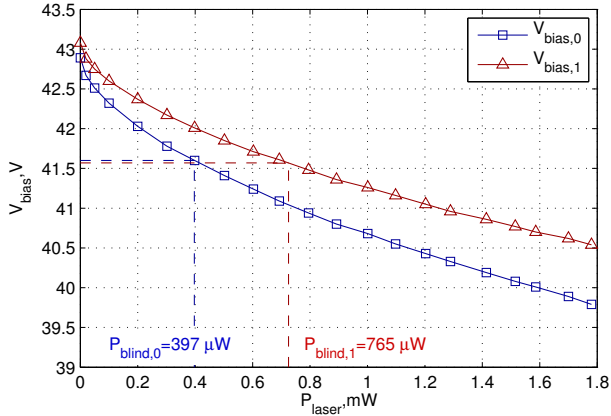}
  \caption{Bias voltage at T1 versus CW laser power for Clavis2. Detector~0 is blind (dark count rate exactly zero) at $P_{\text{laser}} > 397\,\micro\watt$ and detector~1 is blind at $P_{\text{laser}} > 765\,\micro\watt$. QPN 5505 has similar characteristics; due to larger value of $R_{\text{bias}}$, its detector~0 goes blind at $P_{\text{laser}} > 60\,\micro\watt$ and detector~1 goes blind at $P_{\text{laser}} > 85\,\micro\watt$ (see Supplementary information section~II for more details of QPN 5505 blinding).}
  \label{fig:bias-vs-power-cw}
\end{figure}

When eavesdropping, simply applying trigger pulses between the gates populates carrier trap levels in the APD, thus raising the dark count probability and causing a too high quantum bit error rate (QBER). To avoid this Bob's detectors were blinded \cite{makarov2009,makarov2008a}. Then the detectors are insensitive to single photons and have no dark counts. Outside the gates the APD is biased below the breakdown voltage, and the current caused by illuminating the APD is increasing with respect to the incident optical power. A current through the APD will decrease the bias voltage over the APD due to the presence of $R_{\text{bias}}$ (see Fig.~\ref{fig:apds}c) and the internal resistance of the APD. Fig.~\ref{fig:bias-vs-power-cw} shows the bias voltage drop at the point T1 in Clavis2 under CW illumination. 

\begin{figure}[htbp]
  \subfigure[]{\includegraphics[width=8cm]{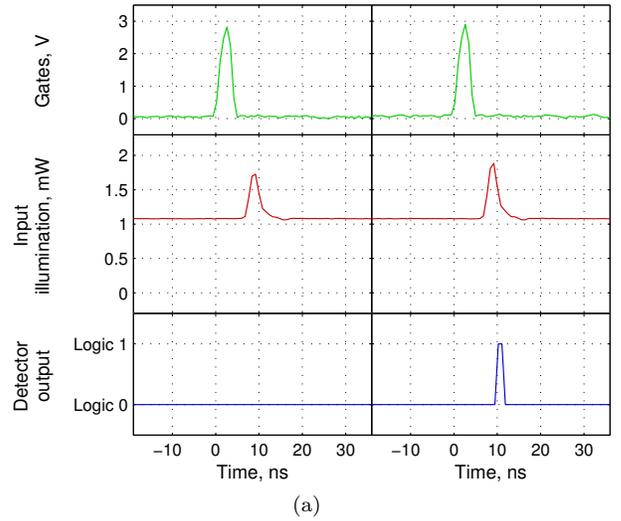}}
  \subfigure[]{\includegraphics[width=\plotwidth]{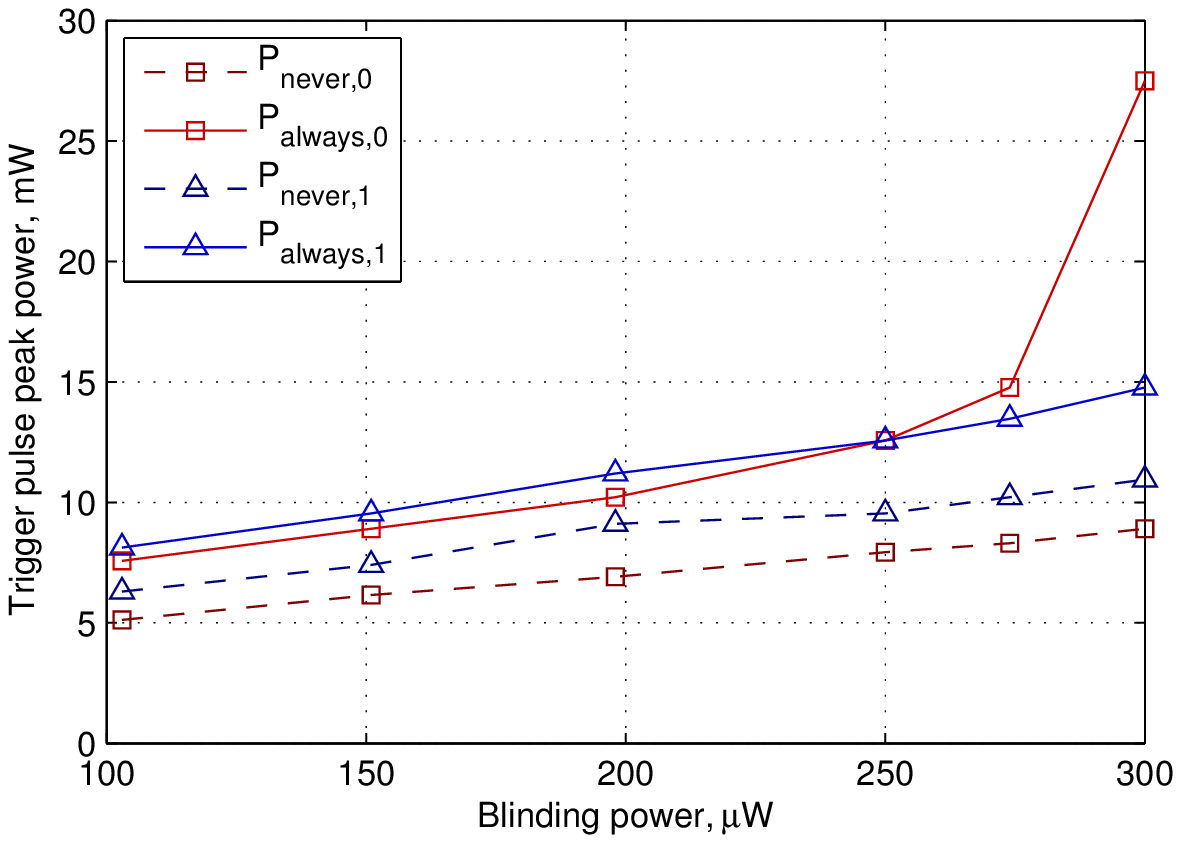}}  
  \caption{Detector control. a) Electrical and optical signal oscillograms when detector~0 in Clavis2 is blinded by $1.08\,\milli\watt$ CW illumination, and controlled by a superimposed $2.5\,\nano\second$ long laser pulse timed slightly behind the gate (see Supplementary information section~III for detailed measurement setup). The superimposed $P_{\text{never},0} = 647\,\micro\watt$ (detector~1: $P_{\text{never},1} = 697\micro\watt$) trigger pulse never causes a detection event while the $P_{\text{always},0} = 808\,\micro\watt$ ($P_{\text{always},1} = 932\,\micro\watt$) trigger pulse always causes a detection event. b) Click thresholds versus the applied CW blinding illumination for the QPN 5505. When the blinding power increases, $P_{\text{always},0}$ diverges, perhaps because the bias voltage is approaching the punch-through voltage of the APD (see Supplementary information section~II).}
  \label{fig:cw-blinding-and-control}
\end{figure}

The blinding is caused by the drop of $V_{\text{bias}}$ such that the APD never operates in the Geiger mode, but rather is a classical photo diode at all times. The voltages $V_{\text{HV},0/1}$ of the high voltage supplies do not change; the entire change of $V_{\text{bias}}$ is due to the resistors $R_{\text{bias}}$. Although shorting this resistor seems like an easy countermeasure, at least for Clavis2 this does not prevent blinding. With higher illumination the electrical power dissipated in the APD generates substantial heat. Raised APD temperature increases its breakdown voltage by about $0.1\,\volt/\celsius$ while $V_{\text{bias}}$ remains constant, which also leads to blinding (at several times higher power level, 4--10$\,\milli\watt$). 

\begin{figure*}[thbp]
\centering
\includegraphics[width=16.5cm]{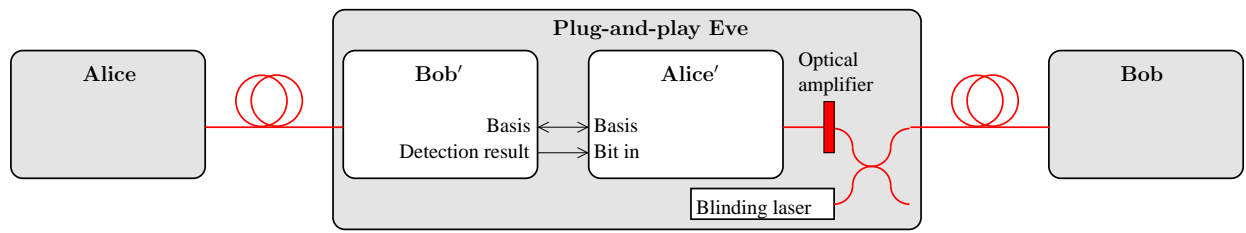}
\caption{Proposed plug-and-play Eve. In the plug-and-play scheme \cite{muller1997} the laser pulses travel from Bob to Alice and back to Bob, passing Bob's interferometer twice. Therefore, polarization drift in the fibre and drift in Bob's interferometer is automatically compensated. Eve consists of copies of Alice (Alice$'$) and Bob (Bob$'$) which share bit and basis settings, a blinding laser, and an optical amplifier used to get the proper trigger pulse power. Due to the plug-and-play principle any environmental perturbations in the fibres Alice--Bob$'$ and Alice$'$--Bob are automatically compensated. See Supplementary information section~IV for a more detailed scheme.}
\label{fig:plug-and-play-eve}
\end{figure*}

To demonstrate detector control in Clavis2, each detector was blinded with $1.08\,\milli\watt$ optical power with a $2.5\,\nano\second$ long trigger pulse superimposed slightly after the gate. Note that a shorter trigger pulse can be timed inside the gate. Fig.~\ref{fig:cw-blinding-and-control}a shows the response of detector~0 in Clavis2 to trigger pulses at the click thresholds. 

Similarly for the QPN 5505 the trigger pulse was timed with its leading edge about $5\,\nano\second$ after the gate. Figure~\ref{fig:cw-blinding-and-control}b shows the click thresholds for the detectors when blinded with 100--300$\,\micro\watt$ CW blinding illumination. In this case, for blinding power levels 100--250$\,\micro\watt$ the detectors remain silent at a power level of $\leq 0.61\cdot P_{\text{always},1}$. 

For both systems the click thresholds fulfill equation~(\ref{eq:attack-requirement}), hence perfect eavesdropping is possible. 

Both systems under investigation operate according to the plug-and-play principle \cite{muller1997} which allows an easily installable plug-and-play eavesdropper (see Fig.~\ref{fig:plug-and-play-eve}). 

A full eavesdropper based on bright-light detector control has previously been implemented and tested under realistic conditions on a $290\,\meter$ experimental entanglement-based QKD system \cite{gerhardt}. Since the attack is clearly implementable, building a full eavesdropper for a commercial cryptosystem would not further expose the problem. A better use of effort is to concentrate on thoroughly closing the vulnerability. An optical power meter at Bob's entrance with a classical threshold seems like an adequate countermeasure to prevent blinding. However, the power meter output should be included into a security proof. Further, the click threshold at the transition between linear and Geiger mode may be very low, allowing practically non-detectable control pulses. How to design hack-proof detectors is unclear to us at this stage, and all future detectors clearly must be tested for side-channels. 

We believe that openly discovering and closing security loopholes is a necessary step towards practical secure QKD, as it has been for multiple security technologies before. For example, RSA public key cryptography has received extensive scrutiny which in the past discovered effective attacks based on implementation loopholes \cite{boneh1999}. In our view quantum hacking is an indication of the mature state of QKD rather than its insecurity. Rather than demonstrating that practical QKD cannot become provably secure \cite{scarani2009a}, our findings clearly show the necessity of investigating the practical security of QKD: Any large loopholes must be eliminated, and remaining imperfections must be incorporated into security proofs. 

Both ID Quantique and MagiQ Technologies have been notified about the loophole prior to this publication. ID Quantique has implemented countermeasures. According to MagiQ Technologies the system QPN 5505 is discontinued; newer models of their system have not been available for our testing. 


\appendix
\section*{Acknowledgements}
This work was supported by the Research Council of Norway (grant no. 180439/V30). Overall cooperation and assistance of the Max Planck Institute for the Science of Light, Erlangen and personally Gerd Leuchs is acknowledged. L.L. and V.M. thank the Group of Applied Physics at the University of Geneva, ID Quantique, and armasuisse Science and Technology for their hospitality, discussions, cooperativeness and loan of equipment. Service of Radiology of the Cantonal Hospital of Geneva is thanked for a quick help in revealing internal layers in a multilayer printed circuit board of a commercial detector.

\section*{Author contributions}
V.M. conceived the idea and planned the study. L.L. and V.M. conducted the Clavis2 experiment with the help of C. Wiechers, D.E. and C. Wittmann. L.L. and V.M. conducted the QPN 5505 experiment. L.L. and J.S. wrote the paper and supplementary information with input from all authors. J.S. and V.M. supervised the project.

\section*{Additional information}
The authors declare no competing financial interests. Supplementary information accompanies this paper.

\end{document}